\documentclass[preprint,flushrt]{aastex}

\newcommand{\nokmsno}{{\rm km~s}\ensuremath{^{-1}}}
\newcommand{\kmsno}{~\nokmsno}
\newcommand{\kms}{\kmsno\ }
\newcommand{\lya}{Ly\ensuremath{\alpha} }
\newcommand{\RXJ}{RX J1230.8+0115}

\begin{document}

\title{ Discovery of a Dwarf Post-Starburst Galaxy Near a High Column Density Local \lya Absorber \footnote{ Based on observations made with the Apache Point 3.5m Telescope, operated by the Astronomical Research Consortium, and the 2.6m Du Pont Telescope of the Las Campanas Observatory, operated by the Carnegie Institution of Washington, D.C. and Pasadena, CA.} }
\author{John T. Stocke, Brian A. Keeney, Kevin M. McLin, and Jessica L. Rosenberg}
\affil{Center for Astrophysics and Space Astronomy, Department of Astrophysical and Planetary Sciences, Box 389, University of Colorado, Boulder, CO 80309}
\author {R.J. Weymann}
\affil{Carnegie Observatories, 813 Santa Barbara St., Pasadena, CA 91101}
\author{Mark L. Giroux}
\affil{Dept. of Physics \& Astronomy, East Tennessee State U., Box 70652, Johnson City, TN 37614}

\shorttitle{Dwarf Post-Starburst Galaxy}
\shortauthors{Stocke et al.}

\begin{abstract}
We report the discovery of a dwarf (M$_B = -13.9$) post-starburst galaxy coincident in recession velocity (within uncertainties) with the highest column density absorber (N$_{\rm H~I} = 10^{15.85}$ cm$^{-2}$ at $cz = 1586$ \kmsno) in the 3C~273 sightline. This galaxy is by far the closest galaxy to this absorber, projected just 71h$^{-1}_{70}$ kpc on the sky from the sightline. The mean properties of the stellar populations in this galaxy are consistent with a massive starburst $\approx$3.5 Gyrs ago, whose attendant supernovae, we argue, could have driven sufficient gas from this galaxy to explain the nearby absorber. Beyond the proximity on the sky and in recession velocity, the further evidence in favor of this conclusion includes both a match in the metallicities of absorber and galaxy, and the fact that the absorber has an overabundance of Si/C, suggesting recent type II supernova enrichment. Thus, this galaxy and its ejecta are the expected intermediate stage in the fading dwarf evolutionary sequence envisioned by Babul \& Rees to explain the abundance of faint blue galaxies at intermediate redshifts. While this one instance of a QSO metal line absorber and a nearby dwarf galaxy is not proof of a general trend, a similar dwarf galaxy would be too faint to be observed by galaxy surveys around more distant metal line absorbers. Thus, we cannot exclude the possibility that dwarf galaxies are primarily responsible for weak (N$_{\rm H~I} = 10^{14-17}$ cm$^{-2}$) metal line absorption systems in general. If a large fraction of the dwarf galaxies expected to exist at high redshift had a similar history (i.e., they had a massive starburst which removed all or most of their gas), these galaxies could account for at least several hundred high-$z$ metal-line absorbers along the line-of-sight to a high-$z$ QSO. The volume filling factor for this gas, however, would be $< 1\%$.
\end{abstract}


\section{Introduction}

The relationship between \lya absorbers and galaxies has been debated in the literature for many years \citep[see e.g.,][]{bahcall96,arons72,weymann81,bergeron91,lanzetta95,steidel98,dave99} and is the source of on-going controversy \citep[see recent volume ``Extragalactic Gas at Low Redshift'';][]{mulchaey02}. The two-point correlation function (TPCF) of absorbers is the only available datum on this relationship at $z\geq$1.6, because at those redshifts even bright galaxies are hard to detect and study \citep[although see][]{adelberger03}. The latest TPCF results continue to draw a distinction between the \lya-only (i.e., apparently metal-free) absorbers and those which also contain metals \citep[e.g., \ion{C}{4} 1548\AA;][]{womble96}. While the metal-line systems show a significant excess power in the TPCF at low velocity differences, the \lya-only absorbers show very little evidence for clustering in velocity \citep[i.e., only a 3-$4\sigma$ excess at $\Delta cz \leq 200$ \kmsno;][]{rauch92,hu95,cristiani97}. A similar result is found at $z\sim0$ for low column density \lya absorbers by \citet[, Papers I-IV hereafter]{penton00a,penton00b,penton02,penton04}. This result is interpreted to mean that metal-line systems are related to galaxies (e.g., bound galactic disks, halos or outflowing winds) while the \lya-only absorbers are more uniformly distributed in space and may not be related to galaxies at all.

At high-$z$, absorption lines of \ion{C}{4} and/or \ion{O}{6} are visible in individual absorption systems down to \ion{H}{1} column densities of N$_{\rm H~I} \approx 10^{14.5}$ cm$^{-2}$ \citep{songaila96,telfer02}. And, based upon analysis techniques that use many, very weak absorption systems in concert \citep[e.g., ][]{cowie98,ellison00,schaye00,telfer02,simcoe04}, metals may be present in a substantial fraction of the still lower column density systems. These TPCF and metal abundance results suggest, but do not prove, that even the uniformly-distributed, lower column density \lya absorbers could contain metals at a level of 10$^{-2}$ to 10$^{-2.5}$ Solar. If this inference is correct, a very early epoch of star formation may have ``enriched'' the intergalactic medium (IGM) more or less uniformly (but see \citealp{simcoe04} for evidence against this conclusion).

At the highest hydrogen column densities ($\geq 10^{17.3}$ cm$^{-2}$) the association between galaxies and absorbers is much clearer. Imaging and spectroscopy of strong \ion{Mg}{2} metal-line systems at moderate $z$ \citep{bergeron91,steidel95,steidel98} have found strong evidence for the association of metal-bearing absorbers with luminous galaxies brighter than $\sim 0.1$L$^*$. In Steidel's (1995, 1998) survey of 55 \ion{Mg}{2} absorbers, 53 have galaxies brighter than $\sim 0.1$L$^*$ within 50h$_{70}^{-1}$ kpc in projection. Since the high equivalent-width \ion{Mg}{2} systems are invariably optically-thick at the \ion{H}{1} Lyman limit, these systems must include both Lyman limit and damped \lya absorption systems. The two absorbers in the \citet{steidel95} sample lacking ``associated'' bright galaxies are damped systems \citep{steidel97} and it is now known that some low-$z$ damped \lya absorbers are associated with dwarf \citep{nestor02} and even low surface brightness (LSB) galaxies \citep{bowen01}. A new blind \ion{H}{1} 21cm survey \citep{rosenberg02} confirms that a very broad range of \ion{H}{1} masses and thus optical and near-IR luminosity galaxies contribute to the damped \lya absorber population \citep{rosenberg03a}. These observations are strongly suggestive that strong \ion{Mg}{2}/Lyman limit absorbers are due to metal-rich gas in bound halos of brighter galaxies ($\geq 0.1$L$^*$), while damped absorbers are due to metal-rich gas in disks of galaxies of all descriptions and luminosities.

Weaker (N$_{\rm H~I}\leq 10^{17}$ cm$^{-2}$) low-ionization, (\ion{Mg}{2}, \ion{Fe}{2}) and high-ionization (\ion{C}{4}, \ion{Si}{4}, and \ion{O}{6}) metal-line absorbers are more numerous and poorly understood \citep[see e.g.,][]{lanzetta95,charlton02,chen00}. One might speculate that these weaker and higher-ionization absorbers are the more diffuse, outer regions of bound halos or, perhaps, the halos or winds associated with less luminous galaxies than found by the \citet{steidel98} survey \citep[see e.g.,][]{lanzetta95}. Particularly enigmatic are the low ionization, weak \ion{Mg}{2} systems \citep{rigby02}. While there is little imaging and spectroscopy of these systems that might reveal ``associated'' galaxies, a photoionization model leads to surprisingly high densities and thus small sizes \citep[$\leq 1$ kpc;][]{charlton02,tripp02,rosenberg03b}, at least along the line-of-sight. Suggestions concerning the enigmatic nature of these very numerous, low-ionization absorbers include extragalactic starforming regions, cool clumps in bound galaxy halos similar to Galactic High Velocity Clouds (HVCs), and cool clumps in unbound galaxy winds. Regardless, if the extremely small inferred cloud depths are indicative of the overall sizes of these weak \ion{Mg}{2} systems, they would be extremely numerous (greatly exceeding the number of bright galaxies) and difficult to understand in any context \citep{charlton02}.

{\it Hubble Space Telescope} (HST) ultraviolet spectroscopy stands to make an essential contribution to our understanding of the relationship between galaxies and the variety of absorbers described above, because some of the HST-discovered absorbers are at sufficiently low redshift that their surroundings can be scrutinized for galaxies in great detail. Statistical evidence for the relationship between low-$z$ absorbers and galaxies is based upon two major HST spectroscopic surveys: (1) the HST QSO Absorption Line Key Project \citep{bahcall93,bahcall96,weymann98,jannuzi98} used the Faint Object Spectrograph (FOS) to study high column density (log N$_{\rm H~I} \geq 14$ cm$^{-2}$) absorbers and (2) Papers~I-IV used the Goddard High Resolution Spectrograph (GHRS) and the Space Telescope Imaging Spectrograph (STIS) to study lower column densities (log N$_{\rm H~I} = 12.5$ to 16 cm$^{-2}$) but over a shorter pathlength ($\Delta z \sim 1$ compared to $\Delta z \sim 3.5$ at $z \leq 0.3$ for the Key Project spectroscopy). The ground-based galaxy survey work of \citet{lanzetta95} based upon the FOS Key Project absorber sample finds that $\sim 30\%$ of these high column density absorbers can be identified with bright galaxies at impact parameters $\leq 160$h$_{70}^{-1}$ kpc. \citet{lanzetta95} and \citet{chen98} extrapolate this result to conclude that all high equivalent width \lya absorbers (N$_{\rm H~I}\geq 10^{14.5}$ cm$^{-2}$) can be identified as the very extended halos of galaxies, although most would be too faint to be detectable at the distance of the Key Project absorbers. On the other hand, investigations of the distribution of galaxies near lower column density (N$_{\rm H~I} \leq 10^{14.5}$ cm$^{-2}$) \lya absorbers, first by \citet[M93 hereafter]{morris93}, then by Paper~III \citep[see also][]{tripp98,impey99} found little evidence for a close association between \lya absorbers and galaxies, although there was some evidence for an association between most ($\sim 78\%$) \lya absorbers and larger-scale galaxy filamentary structures; the remaining 22\% of these absorbers were found in galaxy ``voids'' \citep[Paper~III and][]{mclin02a}. These latter results have suggested to some investigators (including the current authors), that the proximity of high column density absorbers found by \citet{lanzetta95} is not a cause-effect relationship between the absorber and its nearest galaxy, but rather is caused by both galaxies and absorbers being parts of the same large-scale structure filament. This conclusion is supported both by the theoretical N-body simulations \citep[e.g.,][]{dave99} and by the nearest neighbor analysis of Paper~III.

In an effort to utilize the most nearby \lya absorbers to study the absorber/galaxy relationship, we have initiated a deep pencil-beam galaxy survey around sightlines where very nearby ($cz \leq 10,000$ \kmsno) \lya absorbers have been found in the Penton et~al. survey. This sample is ideal for studying the galaxies surrounding low column density \lya -only absorbers and the weak metal-line absorbers because even very faint galaxies can be inventoried. For example, a first result from this survey \citep{mclin02a} finds no galaxy close to several \lya absorbers at $cz \leq 3000$ \kms in galaxy voids, in some cases with absolute mag limits as faint as M$_B = -12.5$. In this Paper we report on a second specific result from our deep optical galaxy survey involving one of the nearest and strongest \lya absorbers in the Penton et~al. study: a low ionization metal bearing absorber at Virgo distance ($cz$=1586 \kms) along the 3C~273 sightline \citep[T2002 hereafter]{bahcall91,morris91,weymann95,sembach01,tripp02}. While our overall survey results will be presented elsewhere (Stocke et al. 2004), here we present the discovery of the closest galaxy-absorber association found by our survey, a dwarf post-starburst galaxy 71h$_{70}^{-1}$ kpc on the sky and $\Delta cz = 50 \pm 50$ \kms away from a high column density (log N$_{\rm H~I} = 15.85$ cm$^{-2}$) \lya absorber. This one absorption system has more \ion{H}{1} column than all of the other \lya absorptions along this sightline, so it is at the very high end of the column density distribution of absorbers studied by our HST and galaxy surveys (Papers~I-IV).
  
This absorber is very similar to the ``weak \ion{Mg}{2}'' absorbers studied by \citet{charlton02}, as described above. T2002 use photoionization models to infer a cloud thickness of 70 pc, quite comparable to the small line-of-sight extents inferred by \citet{rigby02} for weak \ion{Mg}{2} systems. Thus, this absorber and its possibly related galaxy allow us to study weak metal-line absorption systems in close detail. In the next Section we describe the previous work on this sightline and the nearby sightline of \RXJ\ with HST as well as ground-based optical and radio \ion{H}{1} 21cm galaxy survey work in this region. Section 3 describes the discovery and properties of this dwarf galaxy. Section 4 presents a discussion of these results in the context of other work and Section 5 summarizes our conclusions. As with the distances on the sky that we quote above, throughout this paper we will quote distances obtained from redshifts assuming a pure Hubble velocity field in this direction with $H_0$ = 70h$_{70}$ \kms Mpc$^{-1}$. Using the Local Supercluster velocity field model of \citet{tonry00} reduces distance values in this direction by 10\%.

\section{The 1586 \kms Absorber in the Direction of 3C273}

The 3C273 sightline was the first to have intergalactic \lya absorption detected locally using HST \citep{bahcall91,morris91}. It has been extensively studied since then \citep[M93; Papers I-III;][]{weymann98,sembach01} with particular attention to the ``Virgo Cluster'' \lya absorbers at $cz$=1015 and 1586 \kms. Higher signal-to-noise HST/GHRS observations showed that these two \lya absorbers were remarkably well-fit by single Voigt profiles with line widths $b = 69$ and 72 \kms respectively \citep{weymann95}. However, a subsequent far-UV spectrum of 3C~273 taken with the {\it Far Ultraviolet Spectroscopic Explorer} (FUSE) detected 2 and 6 higher-order Lyman lines respectively in these two systems, requiring a substantial downwards revision in the $b$-values to 32 and 16 \kms, respectively. The revised column density for the 1586 \kms absorber is now: log N$_{\rm H~I} = 15.85 \pm 0.1$ cm$^{-2}$. A subsequent STIS echelle spectrum obtained by the STIS instrument team \citep{heap02} detected low-ionization metal species in the 1586 \kms absorber (C~II 1335\AA, S~II 1250\AA\ and Si~1206\AA) and placed sensitive limits on the presence of higher ions, specifically \ion{C}{4}. In addition, we have used an archival GHRS medium resolution spectrum to place a limit on the presence of \ion{Mg}{2} 2798, 2803\AA\ in this absorber of $\leq 45$ m\AA\ ($4\sigma$). A photo-ionization analysis of this absorber by T2002 requires a surprisingly high density of $10^{-2.8}$ cm$^{-3}$, assuming a standard value for the extragalactic ionizing flux at $z \sim 0$ of $1.3 \times 10^{-23}$ ergs s$^{-1}$ cm$^{-2}$ Hz$^{-1}$ \citep{shull99}. A metallicity of 6\%\ Solar and an overabundance of Silicon relative to Carbon of $\sim 0.2$ dex was found, which is strongly suggestive of supernova type II enrichment of this gas. This led T2002 to speculate that this absorber was recently enriched by an outflowing supernova-driven wind. The low metallicity suggests either that the galaxy creating the wind has overall low metallicity or the wind has swept up a large amount of metal-poor gas.

Another bright QSO, \RXJ, is 0.9 degrees away from 3C~273 on the sky (355h$_{70}^{-1}$ kpc at 1586 \kmsno) and also has HST UV spectroscopy available from GHRS and STIS observations \citep[Paper~I and][]{rosenberg03b}. There is a particularly strong \lya absorption line at 1666 \kms with an equivalent width of 389 m\AA, which is almost certainly a blend of two absorbers, both with log N$_{\rm H~I} \approx 16.4$ cm$^{-2}$ given the detection of close pairs of metal lines for a variety of species and ions \citep{rosenberg03b}. In these cases, the line ratios demand a photoionization model, which also leads to small line-of-sight depths through these clouds of $\sim 20$ kpc. The metal abundance of these clouds is consistent with being similar to the 1586 \kms absorber in 3C~273, but with large error bars due to a poorly determined hydrogen column density. The velocity difference between these two absorbers and the 1586 \kms 3C~273 absorber is small enough to suggest that these two absorbers are part of the same structure, which would then be inferred to be $\geq  350$h$_{70}^{-1}$ kpc across the sky \citep{dinshaw97}. However, the coincidence of line pairs in velocity ($\Delta{\rm v} \leq 150$ \kmsno) in QSO spectra can be interpreted either as a physically ``contiguous'' structure or a series of ``correlated'' structures stretching across these sightlines; e.g., metal enriched gas from galaxies aligned along the same large-scale structure filament. In this case ``correlated'' structures are suggested by the much smaller line-of-sight sizes of these two clouds. That is, the 1666 \kms absorbers in the \RXJ\ sightline could be two other small clouds similar to the 1586 \kms absorber studied here, but due to another galaxy aligned along the same large-scale structure filament.

The first systematic attempt to discover galaxies associated with the \lya absorbers towards 3C~273 was conducted by M93, who used deep narrow-band and broad-band imaging in an attempt to detect faint galaxies which might be associated with \lya clouds. They then performed followup multiobject spectroscopy on a ``complete sample of galaxies to ${\rm m_B \approx 19}$'' over an area extending one degree from the QSO in right ascension and 40 arcmin in declination. This corresponds to distances across the sky of $\sim 390$ and 260h$_{70}^{-1}$ kpc respectively at the distance to the absorbers in question. Although the galaxy studied herein is within this sky area and brighter than 19th magnitude, it was not observed by M93 because its image was classified as ``merged'' (S. Morris, private communication). M93 found that, while the 3C~273 absorbers tended to be associated with groups or large-scale structures of galaxies, the \lya absorbers did not cluster with galaxies nearly as strongly as galaxies clustered with each other. The M93 results are consistent with results obtained by other workers using other sightlines \citep[Paper~III,][]{tripp98,impey99}: there is no close relationship between galaxies and low column density \lya absorbers. M93 did not find any new galaxies close to either one of the two Virgo Cluster absorbers.  

A summary of all known galaxies near these Virgo absorbers is given in Table 4 and Figure 7 of Paper~III. For the 1586 \kms absorber there is no previously-known galaxy closer than 250 kpc with $\Delta cz \leq 300$ \kms, to which the absorber could be individually related. Paper~III suggested an alternative model which includes all 7 Virgo \lya absorbers (2 in 3C~273 and 5 in \RXJ) and all 8 known galaxies in this region in a single large-scale structure, which stretches diagonally across these two sightlines for $\geq 20$h$_{70}^{-1}$ Mpc. Including the dwarf we report on here, and a new galaxy discovered recently by the Sloan Digital Sky Survey (SDSS), there are now 6 known galaxies at $cz \approx 1600$ \kms within 1h$_{70}^{-1}$ Mpc of the 3C~273 sightline. The presence of strong absorption systems in these two adjacent sightlines at nearly equal velocity, and containing gas of similar metallicity, is evidence in favor of a ``large-scale structure'' model for these absorbers.

\section{The Dwarf Galaxy Near the 3C~273 1586 \kms Absorber}

Over the past several years we have conducted a deep galaxy redshift survey in regions where low-$z$ \lya absorbers have been discovered with HST. We have used the DuPont 2.6m telescope of the Las Campanas Observatory (LCO) in Chile and the 3.5m WIYN telescope at Kitt Peak National Observatory (KPNO) in Arizona. The LCO observations were taken with the Wide Field CCD multiobject metal slit mask spectrometer (${\rm R} \approx 2000$ for 4000-8000\AA) designed and built by one of us (RJW). A preliminary discussion of the observing setups, data reduction and analysis of both the LCO and WIYN data can be found in \citet{mclin02a}. A more complete discussion of all sightlines observed will be presented later \citep{stocke04}.

Here we present a summary of the results for the 3C~273 sightline, in which spectra for 79 galaxies were obtained using LCO slit masks, only 15 of which had previously been observed by M93. Our spectra of the 15 galaxies previously observed have confirmed the redshifts determined by M93 in all cases. Of the remaining 64, all but 6 were too faint to be selected by M93. Therefore, these new observations extend those of M93 1.5 to 2 magnitudes fainter within 15 arcmin of the sightline.
  
We are confident now that all galaxies with m$_B \leq 19$ and $B \leq 24.2$ mag arcsec$^{-2}$ within 40 arcmin of 3C~273 have had redshifts determined. Our confidence is based primarily upon the work of M93 and our realization that the small incompleteness of the M93 survey was due to a few misclassifications, which have now been rectified; i.e., all objects in this region have been classified by three different automated classification schemes \citep[we have used both the Sextractor \citep{bertin96} and PPP \citep{yee91} image classification algorithms; see][]{mclin02a}.  The magnitude completeness limit of  m$_B \leq 19$ corresponds to M$_B \leq -12.8$ and -12.0 for the more distant and closer 3C~273 Virgo absorbers respectively. Our new survey has extended the M93 work to include $\sim 1/3$ of all galaxies with m$_B \leq 20.6$ within 15 arcmin of this sightline. Thus, this field has been extremely well-surveyed for galaxies to limits comparable to Local Group dwarf spheroidals, and only one galaxy has been found close to the 1586 \kms absorber.

\begin{figure}
\begin{center}
\plotone{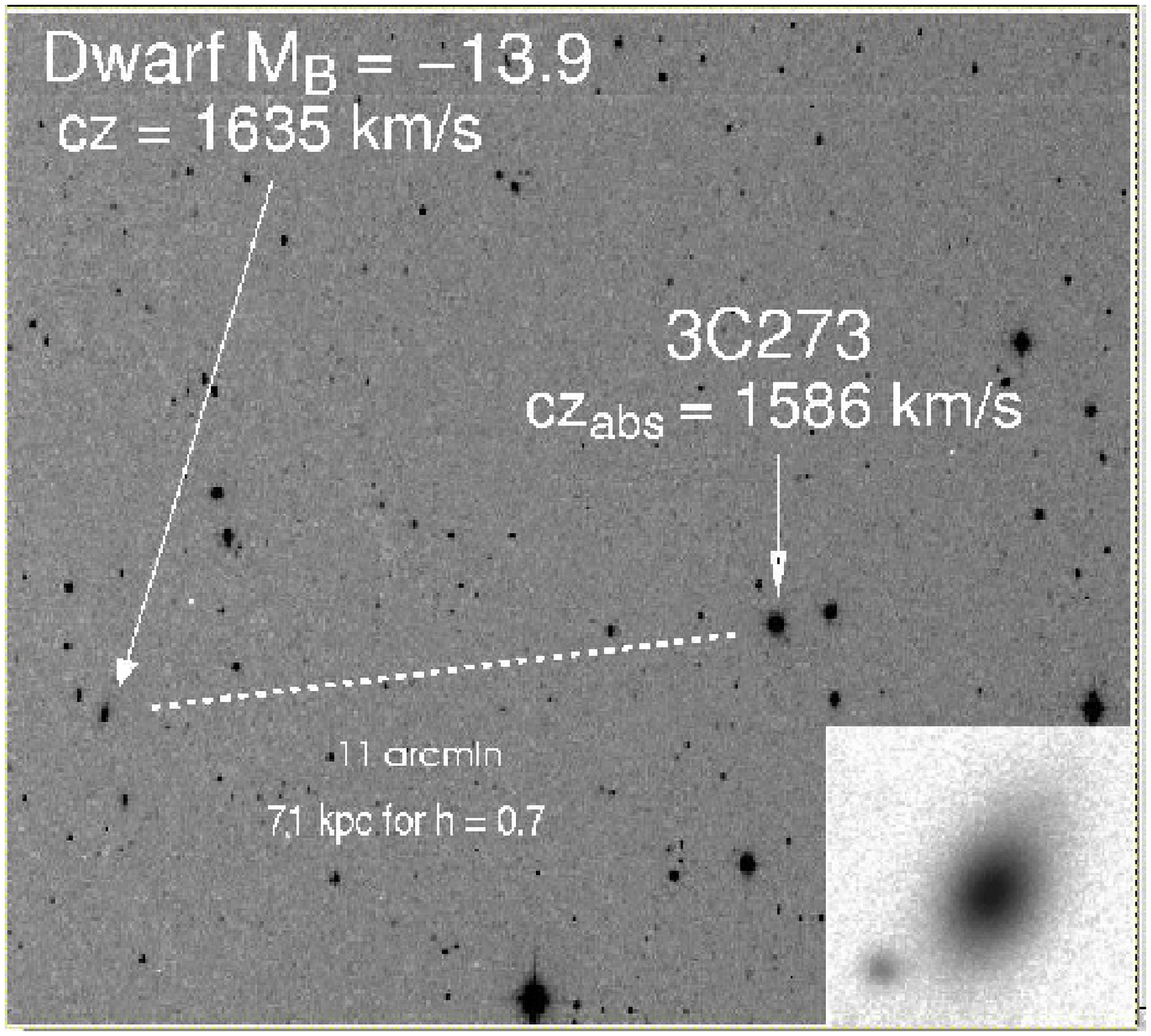}
\end{center}
\caption{Digitized Sky Survey image of the region surrounding 3C~273 (north up, east to the left). Both 3C~273 and the dwarf galaxy have been labeled, along with the heliocentric recession velocities of the dwarf galaxy and the \lya absorber. The nearby QSO discussed in the text, \RXJ, is located 0.9 degrees away from 3C~273 at $PA = 149\arcdeg$. {\it Inset:} A 34 $\arcsec$ square R-band image of the dwarf galaxy taken with the ARC 3.5-m telescope at Apache Point Observatory. The ``companion'' object to the SE is either a foreground star or a globular cluster in the dwarf; it contributes negligibly to the total brightness of the system.}
\label{fig:3C273_dwarf}
\end{figure}

\begin{figure}
\begin{center}
\plotone{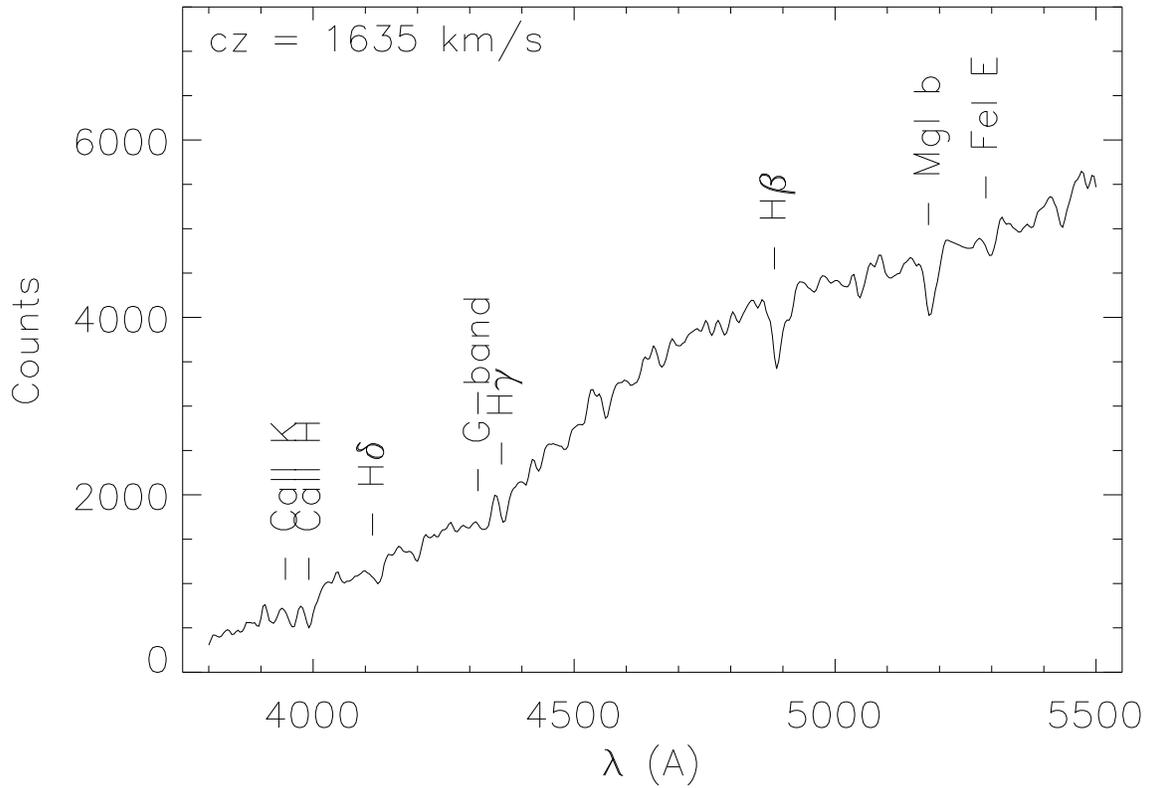}
\end{center}
\caption{Spectrum of the dwarf galaxy with prominent absorption lines marked. We classify this galaxy as a ``post-starburst'' based on the strength of the Balmer lines and lack of emission lines. This spectrum was obtained at the Carnegie Observatory's Las Campanas 2.6-m telescope. Line strengths were measured on an additional spectrum obtained at the ARC 3.5-m telescope at Apache Point Observatory.}
\label{fig:3C273_dwarf_spectrum}
\end{figure}

This galaxy is shown in Figure \ref{fig:3C273_dwarf}; it is a dwarf with  m$_B = 17.9 \pm 0.1$ lying 11 arcmin from the quasar sightline. At the Hubble flow distance to this galaxy, it has a total absolute magnitude, M$_B = -13.9 \pm 0.1$ (0.004L$^*$), and its angular separation from 3C~273 amounts to a projected distance from the  sightline to the galaxy of only 71 h$^{-1}_{70}$ kpc. This is the smallest impact parameter found in our entire survey. This galaxy was observed in two of the Las Campanas masks, as well as in a followup, long-slit spectrum obtained at the Apache Point Observatory (APO) 3.5m using the double imaging spectrograph (DIS). The DIS spectrum has a spectral resolution of $\sim 7$ \AA\ (500 \kmsno) covering the range 3600-9600 \AA. The final redshift we adopt is based upon the 3.5m spectrum because our best LCO spectrum suffers from interference with zero redshift night sky absorption, since it was taken at quarter moon. The velocity obtained from the APO spectrum agrees for the various absorption lines measured to within $\pm 50$ \kms. The velocity determined for the galaxy is $cz =1635 \pm 50$ \kms (heliocentric). A redshift determined from the better of the two LCO spectra is 80 \kms less than the above value. Therefore, the measured radial velocity of this galaxy overlaps the absorber velocity at the $1\sigma$ level. The better of the LCO spectra is shown in Figure \ref{fig:3C273_dwarf_spectrum}.

\begin{figure}
\begin{center}
\plotone{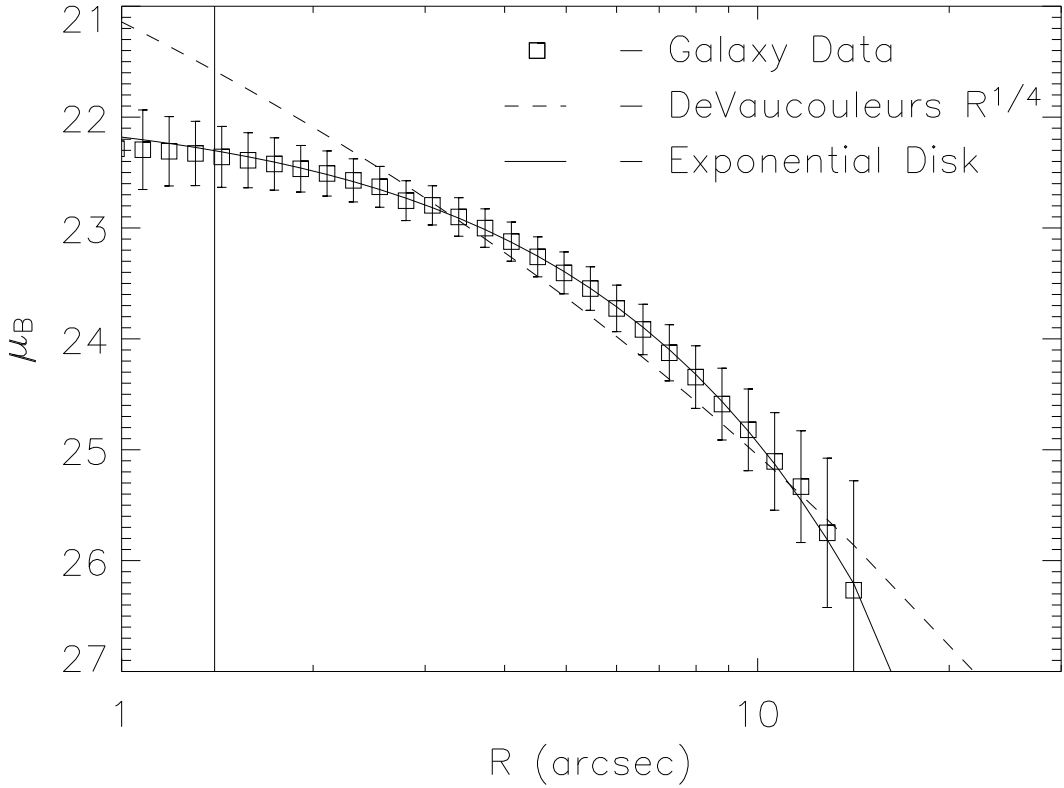}
\end{center}
\caption{Surface brightness profile of the dwarf galaxy (squares) with de Vaucouleurs $R^{1/4}$ (dashed line) and exponential disk (solid line) models overlaid. Like dwarf spheroidals, this galaxy is less centrally concentrated than an elliptical galaxy. The vertical line indicates the seeing (1.4 arcsecs) in the image from which the surface brightness distribution was derived. The small object (star or globular cluster?) just SE of the dwarf in Figure 1 has not been included in the surface brightness profile.}
\label{fig:3C273_dwarf_surf}
\end{figure}

The spectrum can be described as a ``post-starburst'' or $E+A$ galaxy in that both absorption lines of an older or cooler stellar population and hydrogen absorption lines indicating a young, A-F star population are present. No emission lines are visible in the spectrum to a limit of $W_\lambda < 0.6$ \AA, including [OIII] and H$\alpha$ (although not shown in Fig. \ref{fig:3C273_dwarf_spectrum}, the wavelength of H$\alpha$ is present in all three of our spectra). Based upon deep B and R-band images of this galaxy (see inset to Figure \ref{fig:3C273_dwarf}) obtained at the APO 3.5m telescope using the CCD imager ``SPICAM'', this galaxy has a central surface brightness of $B = 21.8$ mag arcsec$^{-2}$, a disk scale length of 0.4h$^{-1}_{70}$ kpc, an overall color of (B--R)=1.3 and isophotes of ellipticity 0.35 with little isophotal irregularities (no significant ellipticity changes or isophotal twists). The seeing during these observations was 1.4 arcsecs. As shown in Figure \ref{fig:3C273_dwarf_surf}, the overall surface brightness brightness profile is much better fit with an exponential disk model than with an r$^{1/4}$-law. Thus, despite its elliptical appearance in Figure \ref{fig:3C273_dwarf}, this galaxy is a spiral, and therefore, similar to Local Group dwarf spheroidals in their surface brightness profiles \citep[e.g.,][]{kormendy86}. Indeed, this object's structural parameters fit nicely onto the dwarf spheroidal structural sequences shown in \citet{kormendy86} and far from the elliptical galaxy sequences. Placing the core radius that we have measured for this galaxy on the dwarf spheroidal sequences in \citet{kormendy86} suggests a velocity dispersion of $\sim 20$ \kmsno. Unfortunately, our LCO and APO spectra were obtained at too low a spectral resolution to confirm this estimate. The evidence that this spiral has been entirely stripped of its gas and dust includes the absence of patchy absorption in the B-band image, the absence of \ion{H}{1} 21cm emission \citep[M$_{\rm H~I} \leq 5 \times 10^6$ M$_{\Sun}$;][]{vangorkom93}, and the absence of H$\alpha$ and [O II] line emission (Table \ref{tbl-lick}). This is consistent with its classification as an $E+A$ galaxy. Given its distance and apparent magnitude, the galaxy is a dwarf with an extrapolated total magnitude of M$_B = -13.9 \pm 0.1$.

Post-starburst or $E+A$ galaxies are quite rare in the local Universe. \citet{zabludoff96} found that only a very small percentage of galaxies in a large field galaxy survey could be described as ``post-starburst'': 0.2\%. However, in a more recent spectroscopic survey of the outskirts of several nearby very rich galaxy clusters, \citet{caldwell97} found that 3-12\% of all cluster ellipticals exhibit post-starburst spectra, although these statistics refer to galaxies considerably brighter than the one we have found here. While Virgo is much less rich than the clusters studied by these authors, we might expect that a few percent of all early-type galaxies in the southern Virgo region would be post-starburst. Thus, an admittedly {\it a posteriori} probability that a post-starburst galaxy is the nearest galaxy to this absorber just by chance is $\sim1\%$. Another {\it a posteriori} probability that supports a physical/causal relationship between this dwarf galaxy and the absorber is that, given that only 6 galaxies total are found in the M93 and current galaxy survey areas within $\pm 500$ \kms of the absorber, the probability that one of these would be found $\leq 70$ kpc from 3C~273 accidentally is $< 5\%$.  So, even if the galaxy filament in this region of sky spans a velocity range of $\sim 1000$ \kmsno, the likelihood is small that a galaxy in this filament would be found just by chance as close to 3C~273 as this dwarf.

The next nearest galaxies to the 3C~273 absorber are UGC 7642 ($cz = 1635$ \kms and m$_B = 15.3$), a starforming dwarf irregular which is 252h$^{-1}_{70}$ kpc away on the sky and UGC 7612 ($cz = 1575$ \kms and m$_B =14.8$), an LSB dwarf Sm galaxy, which is 265h$^{-1}_{70}$ kpc away on the sky. Although these galaxies are $\sim 10$ times more luminous than the dwarf we have found, they are also much farther away from the sightline. Indeed, as we show below, these galaxies have properties which, we believe, that this newly discovered dwarf had several Gyrs ago. There are three other galaxies at comparable recession velocities 483, 743 and 791h$^{-1}_{70}$ kpc away, so there is some evidence for a filament of galaxies in this region at the absorber redshift. So, given that the next two nearest galaxies are at comparable distance from the absorber and that there is some evidence for a galaxy filament (see Paper III), the two alternative hypotheses for the origin of the 1586 \kms absorber are either that it is primarily due to the dwarf we have found or that it is due to the cumulative effect of several nearby galaxies in this filament, including this dwarf
 
\begin{figure}
\begin{center}
\epsscale{1.5}
\plottwo{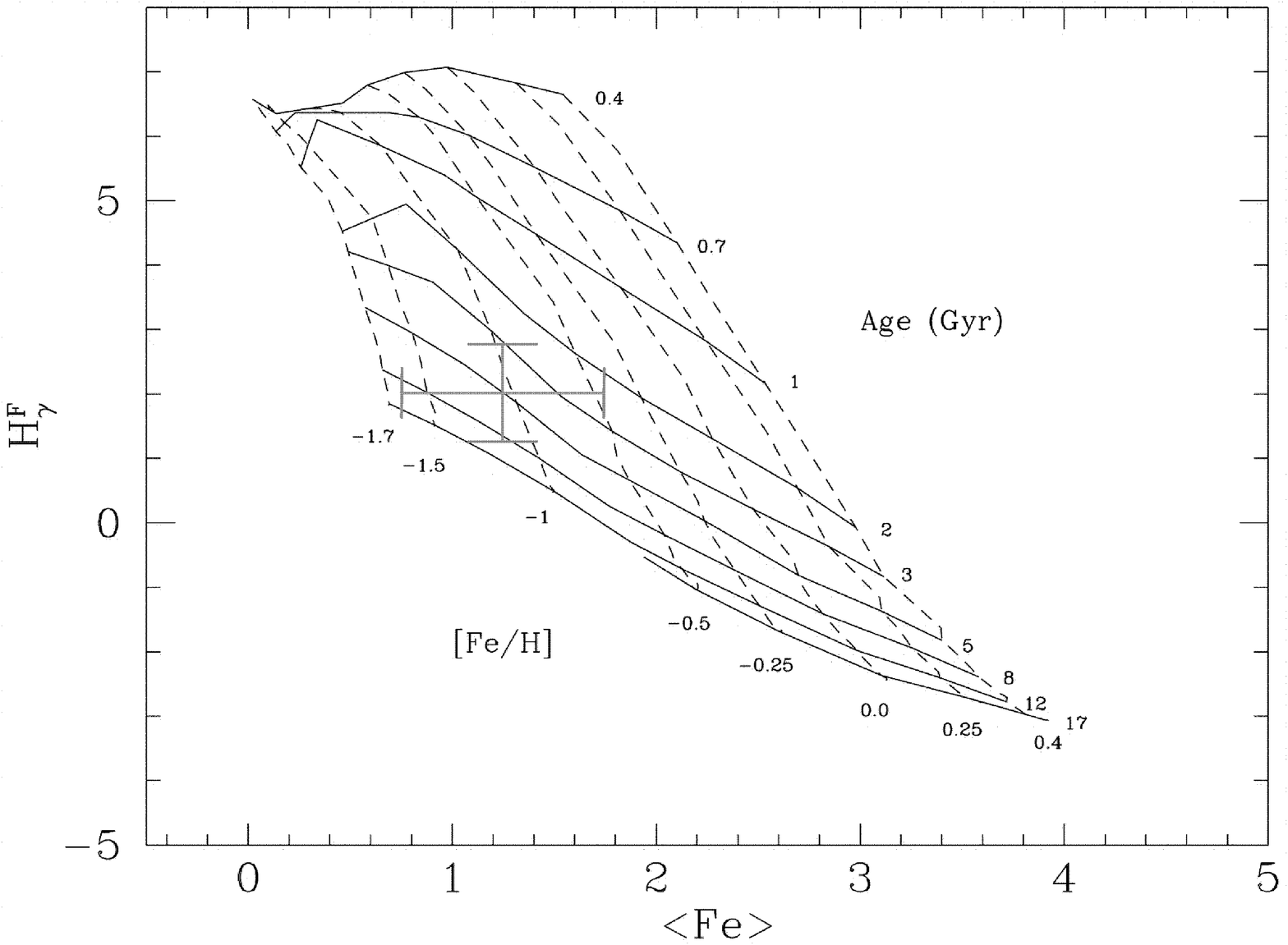}{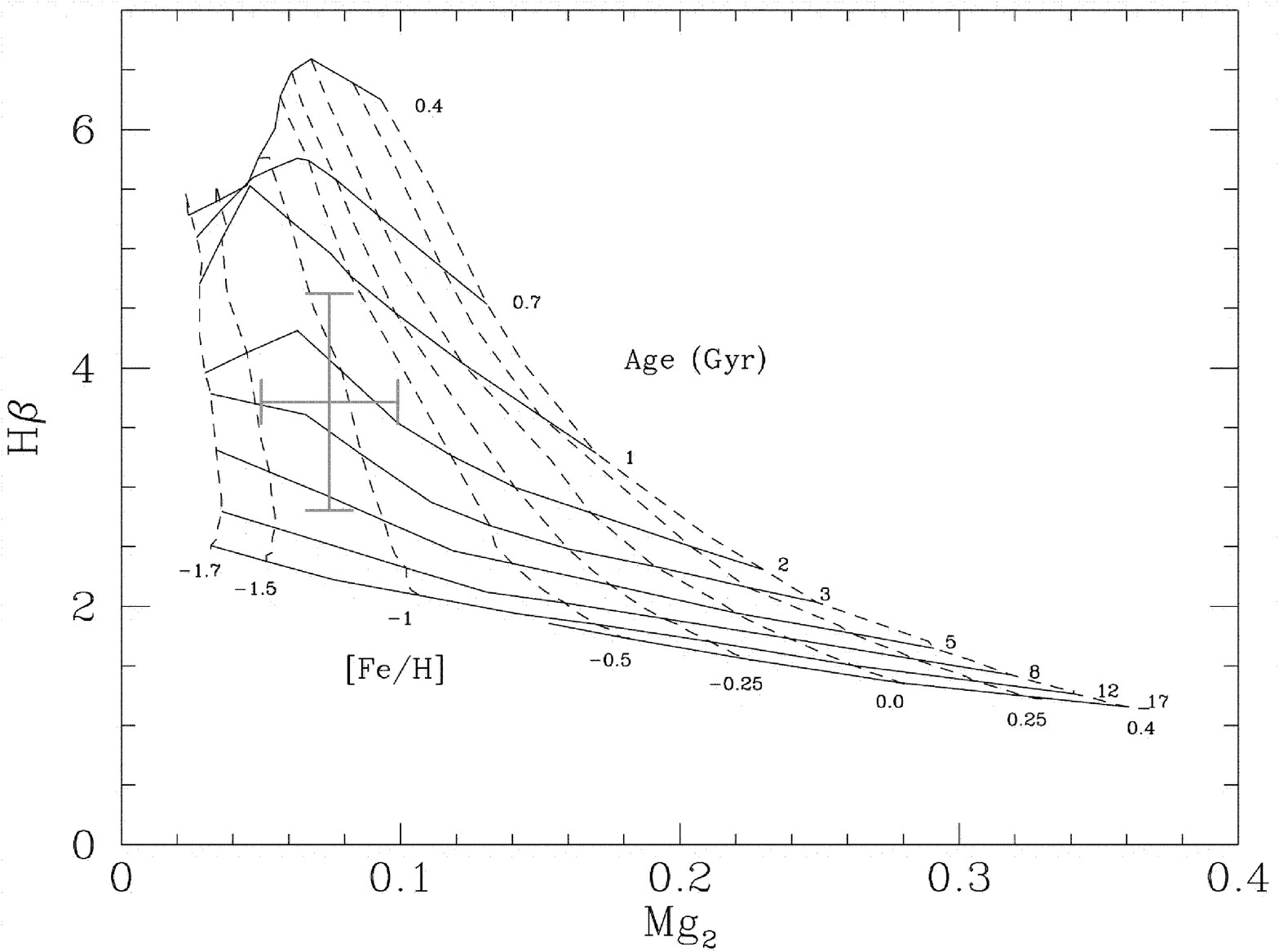}
\end{center}
\caption{Plots of Lick indices based upon the absorption strengths of H$\gamma$ and iron features at 5270 and 5335 \AA\ (top), and H$\beta$ and Mg~I b (bottom), overlaid with population synthesis models from Poggianti et al. (2001). The data point for the dwarf galaxy is shown with error bars. We determine the mean stellar age to be $3.5\pm1.5$ Gyr and the mean metallicity to be ${\rm [Fe/H]} \sim -1.0 \pm 0.5$ dex. This mean metallicity matches the metallicity of the \lya absorber found by Tripp et~al. (2002). Figure courtesy of B. Poggianti.}
\label{fig:3C273_dwarf_lick}
\end{figure}

The APO 3.5m spectrum allows further investigation into the nature of the stellar population(s) in this dwarf. The equivalent widths and line strengths for some of the ``Lick indices'' \citep{worthey94} are shown in Table \ref{tbl-lick}. We also list upper limits on the equivalent widths of the H$\alpha$, [O II] 3727\AA, and [OIII] 5007\AA\ emission lines. Although Figure \ref{fig:3C273_dwarf_spectrum} does not show the H$\alpha$ region, each of our three spectra covered that region and H$\alpha$ was not detected in any of our data to sensitive limits. However, H$\alpha$ does appear to be detected in absorption in our spectrum with equivalent width of $\approx 3$ \AA, so some weak emission may be present which fills in some of the H$\alpha$ absorption line. The strength of H$\delta$ (2 \AA) is large enough to suggest that a single, massive starburst created the F stars which are seen in hydrogen absorption \citep[e.g.,][]{poggianti01,kauffmann03}. However, an H$\delta$ equivalent width of this strength may also be produced from a more extended star formation history \citep[see particularly Figure 3 in][, based upon 10$^5$ galaxies from the SDSS]{kauffmann03}. Therefore, while suggestive of a massive starburst, the Lick indices in Table \ref{tbl-lick} do not require it; while these values constrain the mean characteristics of the stellar population in this dwarf by comparing hydrogen to metal line strengths, they do so regardless of the exact star formation history. Figure \ref{fig:3C273_dwarf_lick} shows the H$\beta$ vs. Mg$_2$ and H$\delta$ vs. $<$Fe$>$ indices as defined by \citet{poggianti01} quoting \citet{worthey94} with a grid of mean population characteristics overplotted. Although the uncertainties are rather large, the two plots in Figure \ref{fig:3C273_dwarf_lick} together require a mean stellar metallicity of -1.0 ($\pm$0.5) dex Solar and a mean stellar age of 3.5 $\pm$ 1.5 Gyrs. If the light from this dwarf is dominated by a single, massive starburst, these values represent the metallicity and the age of the starburst. If the recent star formation history of this dwarf has been more complicated (e.g., an older stellar population from previous star formation accounts for the metal lines), then the location of the Lick index measurements on the population grid in Figure \ref{fig:3C273_dwarf_lick} yields an upper limit on the time in the past when a less massive starburst occurred.

Multiple epochs of star formation found in many local group dwarf spheroidals \citep{mateo98} are indirect evidence in favor of the latter interpretation. These two possible histories of this dwarf can be discriminated by obtaining a near-UV image with HST/STIS. If the luminosity of this galaxy is due primarily to a single starburst 3.5 Gyrs ago, the near-UV flux should be quite faint, 2.5-3 mags fainter than the B-band flux. If a more extensive star formation history is correct, numerous blue, horizontal branch stars will be present making the near-UV flux substantially brighter \citep{brown97,tantalo96}. Since the evidence that some dwarf spheroidals have had star formation histories that include several star forming episodes, a single, massive starburst could be unusual. However, this galaxy is one of the few dwarf spheroidals known that is not close to a much more massive galaxy, whose tidal forces could have stimulated several star forming episodes.

\section{Discussion}

\subsection{Relationship Between the 3C~273 Absorber and the Nearby Dwarf Galaxy}

A summary of the important measured properties of the 3C~273 1586 \kms absorber and the ``post-starburst'' dwarf galaxy are shown in Table \ref{tbl-compare}. These reference values include the observed column density (N$_{\rm H~I}$), and the density (n) and ionization parameter (U) from the photoionization model of T2002 as well as dwarf galaxy parameters from Section 3 including the 3C~273 sightline impact parameter (b). Briefly, the correspondences which strongly suggest a physical/causal relationship between this galaxy and the absorbing gas are:
\begin{enumerate}
\item The proximity on the sky and in recession velocity. Within the errors the absorber and galaxy redshifts match. The galaxy type is unusual if the absorber and galaxy are close just by chance.
\item Within rather large errors the metallicities of the galaxy and the absorber match.
\item The super-Solar [Si/C] ratio in the absorber suggestive of recent supernova type II enrichment and the evidence for a recent starburst in the dwarf (Balmer absorption but no emission lines) seem to link these two together causally.
\end{enumerate}

While these comparisons are strongly suggestive, the metal enrichment of such a large region of the IGM by such a small galaxy may seem extreme and deserves more detailed scrutiny. First, while the evidence is inconclusive that this galaxy's stellar population is due to a single, massive starburst $\sim 3.5$ Gyrs ago, there is strong spectroscopic evidence for recent star formation and a current absence of \ion{H}{1} (see Table \ref{tbl-compare}). So, although the number of recent supernovae and the amount of gas expelled from this galaxy are not precisely known, nevertheless, there is indirect evidence that a supernova driven wind did expel most of this galaxy's gas and dust into the IGM. The absence of photometric peculiarities in the surface brightness profile shown in Figures \ref{fig:3C273_dwarf} and \ref{fig:3C273_dwarf_surf} argue against the gas being stripped by a recent interaction with another galaxy; nor is a nearby galaxy seen. And although this galaxy is on the southern outskirts of the Virgo Cluster, its distance, 3.3h$_{70}^{-1}$ Mpc away on the sky, places it well outside the region where ram pressure stripping by intra-cluster gas would be effective. The expected low escape velocity for this dwarf ($\leq 50$ \kmsno; see below) makes gas ejection into the surrounding IGM seem more plausible than if this galaxy was much more massive. Indeed, the combination of a post-starburst dwarf and a plausibly associated \lya absorber is among the strongest pieces of observational evidence showing that starburst driven winds actually escape from galaxies \citep[see e.g.,][]{martin03}.

To pursue the dwarf galaxy wind scenario, for the moment we will assume that the galaxy characteristics we see are due to a single, massive starburst 3.5 Gyrs ago. In this case, the absolute B magnitude of the dwarf is almost entirely due to main sequence F stars and the implied current luminosity of the starburst is nearly $\sim 10^8$ Solar luminosities (i.e., the fading starburst accounts for virtually all of the galaxy's observed B-band luminosity). Bruzual and Charlot's (1993) stellar synthesis models of instantaneous starbursts suggest that this dwarf would have been $\geq 10$ times brighter in B-band within the first $10^8$ yrs after the starburst (i.e., M$_B \approx -16.5$) and will fade to be $\sim 10$ times fainter in the next few billion yrs (i.e., M$_B \approx -11.5$). This final luminosity and the dwarf's structural properties would then be similar to those observed for the local group dwarfs Sculptor, Leo I \& II and Sagittarius. The original \ion{H}{1} content of the galaxy can be estimated based upon the galaxy's brightness during its starburst phase (its most detectable phase). From the results of a blind \ion{H}{1} survey with the Arecibo dish, \citet{rosenberg02} find that galaxies with mags comparable to this dwarf when it was ``bursting'' have a median M$_{\rm H~I} \approx 5 \times 10^8$ M$_{\Sun}$, with an expected range of $10^{8-9}$ M$_{\Sun}$ in \ion{H}{1}. As \citet{rosenberg02} is an \ion{H}{1} selected galaxy survey, these numbers are biased towards higher \ion{H}{1} masses but an original \ion{H}{1} content of a few x 10$^8$ M$_{\Sun}$ seems quite plausible for this dwarf. This is enough mass to account for the nearby 3C~273 absorber.

Assuming that the 1586 \kms absorber is part of a complete, spherical shell surrounding this galaxy, it has a thickness and density inferred from the absorption line data of 70 parsecs and 1.4 x 10$^{-3}$ cm$^{-3}$ (T2002) respectively and a minimum radius of 70 kpc, given by the observed impact parameter (see Table \ref{tbl-compare}). This geometry requires a total mass of $\sim 10^8$ M$_{\Sun}$, which should be considered as an upper limit to the mass the galaxy must supply because the shell might not be entirely intact and the superwind from the galaxy may have swept up some ambient gas on its way out to the absorber location. Thus, if a large fraction of the \ion{H}{1} in this galaxy were blown out by supernovae, it could account for the 3C~273 absorption that we see. Using a 70 kpc radius assumes that we are viewing the very edge of the shell towards 3C~273. This is a reasonable assumption since we see only one absorber, although a single absorber could also be produced by a larger shell of gas which has only a partial covering factor, requiring somewhat larger mean velocities.

In two other Penton et~al. survey (Papers I \& IV) sightlines (\RXJ\ and PG 1211+143), \lya plus metal-line absorbers with comparable \ion{H}{1} column density to this one have at least two components \citep{tumlinson04, rosenberg03b}. By the current interpretation, pairs of lines are suggestive of sightlines which lie along a cord of the superwind's shell. The \RXJ\ absorber pair has an observed radial velocity difference of 36 \kms \citep{rosenberg03b}. This radial velocity difference implies a similar outflow velocity to that required here assuming a spherical shell-like geometry. If the dwarf superwind hypothesis is correct, we would expect to find a small galaxy near to the \RXJ\ 1666 \kms absorbers, similar to the dwarf studied here. While a galaxy redshift survey is in progress near \RXJ, we have not yet found a galaxy comparably close to \RXJ \footnote{While the nearest known galaxies to the \RXJ\ absorbers, A1228+0116 \& NGC~4517A (Paper III), are 118h$^{-1}_{70}$ and 176h$^{-1}_{70}$ kpc away from the \RXJ\ sightline at 1666 \kmsno, they have large velocity differences ($+625$ \kms and $-560$ \kmsno) from the absorber, and so are not so obviously related to these absorbers as the dwarf discussed here is plausibly related to the 3C~273 absorber.}. And, even at the modest distances (for QSO absorbers) at which the two metal-line absorbers in PG~1211+143 are found ($cz \sim$ 14, 000 and 19,000 \kms), a dwarf galaxy like the present case would have m$_B >$ 22, well below the sensitivity of current surveys \citep[e.g.,][]{stocke04}. However, for each PG~1211+143 metal-line absorber, a more luminous galaxy is found projected $\sim 100$h$^{-1}_{70}$ kpc at comparable recession velocity \citep{tumlinson04}. Thus, for the case of PG~1211+143, and for numerous other metal-line absorbers (including those at $z \sim$ 3; \citet{adelberger03}), the presence of luminous galaxies does not preclude, and is, in fact, suggestive of, the presence of fainter galaxies closer to the absorber, as we have found here.

Could this little galaxy's supernovae blow this amount of gas out of the galaxy to the location seen? Taking 3.5 $\pm$ 1.5 Gyrs ago to be the time of the ejection of the gaseous shell, a mean space velocity of $\geq$20 $\pm$10 \kms is required to reach a radius of $\geq$70h$_{70}^{-1}$ kpc. This velocity seems entirely plausible, if not smaller than expected for such an outflow given the large gas velocities and temperatures seen close to the star forming region \citep[e.g.,][] {martin03, heckman01, pettini01}. Given that the absorber and galaxy radial velocities match to within the errors, a modest outflow velocity of this order is entirely consistent with the observations. Were there enough supernovae in the starburst to create this outflow velocity? In order to estimate the number of supernovae in the starburst, we assume a mass-in-stars to light ratio of $M_{stars}/L_B$ = 1-5. We obtained these values from the Bruzual \& Charlot (1993) instantaneous burst model for 5\% stellar metallicity at a time after the burst of $3.5 \pm 1.5$ Gyrs. These mass-to-light ratios yield a stellar mass in the starburst of 1-$5 \times 10^{8}$ M$_{\Sun}$. We would expect that the total dynamical mass of this galaxy is a few times higher (i.e., $M/L_B \approx 10$, comparable to the local group dwarf spheroidals; Mateo 1988). Assuming that mass follows light, the surface mass profile (Figure 3) and ellipticity (0.35; Section 3) yield an escape velocity of $\sim 50$ \kms at one disk scalelength (0.4h$^{-1}_{70}$ kpc). This is consistent with the estimate from different considerations above. If the starburst created $\geq 10^{8}$ M$_{\Sun}$ of stars, a Salpeter initial mass function predicts $\geq$ 7 x 10$^5$ stars with $>$ 8 M$_{\Sun}$, which then create supernovae. Assuming 10$^{51}$ ergs of energy per supernova, converted to bulk kinetic energy at 3-30\% efficiency \citep{koo92a,koo92b,cioffi91}, there was $\geq$ 2 $\times$ 10$^{55-56}$ ergs available to produce a superwind. After overcoming the shallow potential well of this dwarf, this amount of energy is capable of accelerating several $\times$ 10$^8$ M$_{\Sun}$ of gas to a mean velocity of 20-30 \kms. Thus, if a single massive starburst created most of the stars we see in this galaxy today, the subsequent supernova explosions seem entirely capable of producing a superwind which could account for the 3C~273 absorber.

However, a starburst of $\sim 1$ M$_{\Sun}$ yr$^{-1}$ for 10$^8$ yrs is quite large among dwarfs \citep{martin03} and the star formation history of this dwarf may have been more complicated; i.e., the observed stellar population is a combination of a young starburst population and an older population(s) due to previous burst(s) or to a steadily declining star formation rate. A variety of more complicated star formation histories is observed for the Local Group dwarf elliptical and spheroidal galaxies \citep{mateo98}. In this case, to satisfy the constraints provided by the Lick indices in Table \ref{tbl-lick}, the most recent starburst must have been less massive and more recent than in the single burst model above. Even if the most recent starburst in this dwarf was $\sim0.3$ M$_{\Sun}$ yr$^{-1}$ and 2 Gyrs ago, it would still be capable of producing the required outflow. A smaller, more recent starburst than this would not be capable of accelerating the required $\sim 10^{8}$ M$_{\Sun}$ of gas to $\geq 70$h$_{70}^{-1}$ kpc in time to create the absorber we see. In the case of an $\sim 0.3$ M$_{\Sun}$ yr$^{-1}$ starburst, the brightness of this dwarf while bursting would be 3-10 times brighter than what we see today but it will not fade significantly in the future, as its current brightness is already partially due to an older stellar population. In this case, the eventual end-point of this system will be more similar to the dwarf elliptical, NGC 185. Since we based our estimate of the original \ion{H}{1} content of this dwarf (M$_{\rm H~I} = 5 \times 10^8$ M$_{\Sun}$) on its luminosity while ``bursting'', this estimate should be revised downwards by a factor of a few if a more complicated star formation history is correct. However, this amount of gas is still sufficient to produce the 1586 \kms absorber, especially since some primordial gas may be swept up as the wind escapes into the IGM. While a more ancient starburst ($> 5$ Gyrs ago) could in principle account for this absorber as well, it would not be consistent with the Lick indices we have measured for this dwarf. In summary, as long as this dwarf galaxy had a significant ($\gtrsim 0.3$ M$_{\Sun}$ yr$^{-1}$ for $\sim 10^8$ yrs) starburst more than $\sim 2$ Gyrs ago, it can plausibly account for the metal-line absorber in the 3C~273 sightline 70h$_{70}^{-1}$ kpc away. Such a starburst is strongly suggested, but not unambiguously demanded, by the Lick absorption line indices, and the absence of \ion{H}{1} 21cm and optical emission lines that we find for this dwarf.

As described in detail in \citet{rosenberg03b}, nature has provided us with two sightline probes through the $cz = 1500$-1750 \kms galaxy and gaseous filament in this region. \RXJ\ is 0.9$\arcdeg$ (355h$^{-1}_{70}$ kpc at $cz = 1600$ \kms) away from 3C~273 and both sightlines possess N$_{\rm H~I} \approx 10^{16}$ cm$^{-2}$ \lya absorbers at $cz \sim 1600$ \kms with quite similar properties, including metal abundance. This suggests that the covering factor of metal enriched gas in this filament is approximately unity, which in turn requires a few (3-6) starburst galaxy outflows scattered through the filament to produce a high covering factor for metal-enriched gas. This is a reasonable number of dwarf galaxies in this region, since six dwarf galaxies are already known to be present. If a seventh dwarf galaxy is discovered within $\sim 100$h$^{-1}_{70}$ kpc of the \RXJ\ sightline at $cz = 1600$-1750 \kmsno, this discovery would solidify the dwarf starburst + superwind hypothesis for these absorbers. However, a deep (e.g., m$_{\rm B} \leq 19$) galaxy survey in the \RXJ\ region has not yet been completed.

Also, this dwarf could be an example of a ``fading blue dwarf'' as envisioned by \citet{babul92} to explain the large numbers of faint blue galaxies found in deep optical images \citep[e.g.,][]{tyson88}. If no further star formation occurs in this dwarf, several Gyrs from now this galaxy will fade to the brightness of a dwarf spheroidal, which it already resembles in its surface brightness distribution. This dwarf and its associated nearby absorption line system are the expected intermediate stage between a more luminous blue dwarf actively forming stars and the abundant dwarf spheroidal galaxies, which have no current star formation. The superwind scenario of \citet{babul92} predicts just the type of association we have found between a fading dwarf and a QSO absorber. If we were able to view another dwarf with properties like this one had $\sim$3.5 Gyrs ago (and thus at an observed $z=0.3$-0.5), it would have m$_B = 24$-25, the magnitudes and redshifts \citep[see e.g.,][]{koo95} where the faint blue galaxies are becoming prominent in deep images.

\subsection{The QSO Absorption Line Context}

In this paper we have proposed that a starburst $\sim 3.5$ Gyrs ago created an outflowing superwind which escaped from a dwarf galaxy, expanded $\sim 100$ kpc into the IGM, enriching the IGM in metals and energy, and creating a \lya + metal absorber that we observe in the UV spectrum of 3C~273. While this individual case may be unique, these circumstances could be much more general because the 3C~273 1586 \kms absorber is the closest example of an IGM metal line system currently known. And, if this absorber were much more distant, the dwarf galaxy we propose to be the source for the absorbing gas would be too faint to be observed by current galaxy surveys near QSO absorbers \citep[e.g.,][]{morris93, penton02, stocke04}. Thus, the frequency of dwarfs near QSO metal line absorbers is not known currently. Therefore, we now consider how frequently this circumstance {\it could} occur and thus how many QSO metal absorption line systems {\it could} be attributed to starbursting dwarf galaxies?

This possibility has been explored theoretically by a number of authors \citep[e.g.,][]{maclow99,ferrera00} who found that, while a complete ``blow away'' of gas from a dwarf was possible only if M$_{\rm H~I} < 10^7$ M$_{\Sun}$, dwarfs could generate powerful winds with more than escape speed through the combined action of their supernovae. The models of \citet{ferrera00} predict that dwarf galaxies with M$_{\rm H~I}\approx 10^8$ M$_{\Sun}$ are the dominant contributor of gas and metals from galaxies into the IGM. Significant observational evidence \citep{heckman01, martin03} exists that there is little correlation between velocity or gas temperature measured for superwinds and galaxy luminosity suggesting, but not proving, that dwarf superwinds escape but winds from more luminous galaxies do not. \citet{ott01} have studied a possible nearby example of this phenomenon, the dwarf galaxy Holmberg I in the nearby M81/82 group.  So this hypothesis is supported both theoretically and observationally. Supernovae in more massive galaxies produce winds which do not escape but instead form ``galactic fountains'' \citep{ferrera00}. Based upon these theoretical results and the present example, we suggest that the \ion{H}{1} column density range where dwarf superwinds may be the dominant source of metal-enriched gas is 10$^{14 - 17}$ cm$^{-2}$.

At higher column densities (N$_{\rm H~I} \geq 10^{17.3}$ cm$^{-2}$) ground-based observations \citep[e.g.,][]{bergeron91,steidel95,steidel98} strongly suggest that \lya + metal-line absorbers are due to bound halos of luminous galaxies. The standard schematic geometry \citep{steidel98}, suggested by the number of strong \ion{Mg}{2}/Lyman limit systems ($\sim 1$ per unit redshift), has N$_{\rm H~I} \geq 10^{17.3}$ cm$^{-2}$ absorbers arising in 50h$^{-1}_{70}$ kpc halos around bright galaxies; the small percentage of Ly limit systems that are also damped-\lya systems have numbers that are consistent with arising in significantly smaller regions that are the size of galaxy disks. The galaxy survey work of \citet{steidel95,steidel98} near these strong \ion{Mg}{2} absorbers at $z < 1$ solidifies this picture, in that he finds luminous (${\rm L} > 0.1$L$^*$) galaxies close to the sightline at the redshift of the \ion{Mg}{2} absorber in 53 of 55 cases \citep[the remaining two cases are both damped \lya systems;][]{steidel97}. The low-z work on damped systems by \citet{turnshek02}, \citet{bowen01}, and others have solidified the hypothesis that damped \lya systems arise in the gaseous disks of galaxies of a wide variety of types \citep[including low surface brightness (LSB) galaxies and dwarfs;][]{bowen01,rosenberg02}. Thus, the two undetected galaxies associated with damped \lya absorbers in Steidel's survey are explicable based upon these more recent results; i.e., these are probably due to dwarfs or LSBs undetectable at intermediate redshifts. \citet[with details found in \citealt{mclin02b}]{mclin98} showed that Steidel's general galaxy survey result rules out dwarf galaxies being responsible for the Lyman limit systems which are not damped \lya absorbers. This is because Steidel's success rate of identifying luminous galaxies at the absorber redshift is much higher than would be expected if low luminosity dwarfs contribute to forming these systems. Specifically, \citet{mclin98} showed that for several plausible choices of faint end slopes to the galaxy luminosity function, a gaseous dwarf halo with radius as extrapolated from the halo sizes found by \citet[i.e., 20-40 h$^{-1}_{70}$ kpc]{steidel98} was inconsistent with always finding a bright galaxy associated with the absorber. \citet{mclin98} concluded that the absence of strong \ion{Mg}{2} systems near dwarfs could be understood if dwarfs could not retain large gaseous halos, but instead developed unbound winds. While these winds are predicted to be quite extensive, they rapidly become optically-thin at the Lyman limit and therefore were unlikely to be strong \ion{Mg}{2}/Lyman limit systems. From completely different considerations \citet{mo96} also found that extensive gaseous halos were unlikely to surround dwarf galaxies. So, from the above observations and interpretations, we can understand the threshold at ${\rm L} = 0.1$L$^*$ (v$_{escape} = 100$-150 \kms), below which Steidel finds no galaxies associated with Lyman limit absorbers. We suggest that this limit is due to the galaxy gravitational potential above which supernovae-driven superwinds do not escape from a galaxy, but instead create a ``galactic fountain'' and so form an optically-thick gaseous halo. A nearby example of a bound halo is the 3C~232/NGC~3067 absorption system, for which HST/GHRS spectra verify the strong \ion{Mg}{2}/Lyman limit system \citep{tumlinson99} and an \ion{H}{1} 21cm emission image of the gas verifies that the QSO shines through the galaxy's extended gaseous halo \citep{carilli92}. These results are consistent with the \citet{maclow99} and \citet{ferrera00} simulations.

While these results suggest that weaker metal line systems might be due to the more tenuous, outer halos of luminous galaxies, these systems are much more numerous \citep{young82,steidel98} and luminous galaxies have not generally been found near these weaker absorbers \citep[although see \citet{adelberger03} as discussed below]{rigby02}. The number of these weak metal line systems is large, and if extrapolations made from current data are correct, may include the bulk, if not all, of the \lya forest \citep{cowie98,schaye00,ellison00}. If only those \ion{C}{4} absorbers associated with \lya absorption at N$_{\rm H~I}\geq 10^{15.5}$ cm$^{-2}$ are considered, their numbers are $\sim 10$ times greater than the strong \ion{Mg}{2} absorbers and thus would require high covering factor halos of radius 100 h$^{-1}_{70}$ kpc around galaxies \citep{steidel98}. More sensitive observations have found \ion{C}{4} associated with absorbers down to N$_{\rm H~I}\geq 10^{14.5}$ cm$^{-2}$, with suggestions that \ion{C}{4} (and \ion{O}{6}) absorptions persist to even lower column densities based upon pixel optical depth techniques using many, lower column density absorbers \citep{cowie98,aguirre02}. The ensemble metallicity of these weak \ion{C}{4} and \ion{O}{6} absorbers is in the range of 0.3-1\% Solar. To associate these even lower column density metal systems with bound halos seems untenable given their huge numbers, since current observations are consistent with a large fraction, if not the entire \lya forest, having a metallicity with Z $\geq 0.3\%$ Solar. At N$_{\rm H~I} \leq 10^{14.5}$ cm$^{-2}$ current data are consistent either with a metallicity ``floor'' throughout the Universe of 0.3\% Solar or there could be a mixture of metal-enriched and metal-free absorbers in the \lya forest, with the fraction of metal-free absorbers increasing with decreasing hydrogen column density \citep{simcoe04}. Either way, any model for metal line absorbers must account for dN/dz $\geq 100$. In principle, dwarf galaxy starburst superwinds could account for a number density this large, even assuming a ``standard'' \citet{schechter76} faint-end slope to the local luminosity function of galaxies of $\alpha = -1.25$. Assuming that all dwarf spheroidal galaxies went through this starburst phase in the past and that all low luminosity spirals, irregulars, blue cluster dwarfs, etc. are capable of producing such winds, we integrate the luminosity function from 0.1L$^*$ (above which galaxy halos are bound) to 0.001L$^*$ (the luminosity of the Local Group dwarf spheroidals). Regardless of luminosity we assign a cross-sectional radius to the metal-enriched region of 75h$^{-1}_{70}$ kpc, based upon the dwarf galaxy/\lya absorber connection we report in this paper. If each and every low luminosity galaxy produced a sphere of metal-enriched gas of high covering factor over a region of radius 75h$^{-1}_{70}$ kpc, this process can produce $\sim 200$ metal-line systems per unit redshift at $z = 2$ (n.b., the assumption made here of a constant cross sectional area per galaxy regardless of luminosity differs from the standard assumption of an increasing halo size for more luminous galaxies, from which \citet{steidel98} derives $\sim 100$h$^{-1}_{70}$ kpc \ion{C}{4} halos for L$^*$ galaxies). But there may have been even more small galaxies at high-$z$ than in the current epoch, because these objects merged to create the higher luminosity galaxies of today. A conservative estimate of the number of these small objects is a factor of 3 times higher than the estimate above \citep{gnedin00,ricotti02,mo02}, so that the number density of metal-line systems inferred to be produced in this way could be $> 500$ per unit redshift. Thus, dwarf galaxy winds could, in principle, account for all metal-line absorbers down to N$_{\rm H~I} \gtrsim 10^{14}$ cm$^{-2}$ \citep[see, e.g.,][]{bechtold94}. Because these dwarfs fade as a natural part of their evolution as envisioned by \citet{babul92}, they would be very difficult to detect at all but the lowest redshifts. We note that these galaxies are remarkably similar to the ``Cheshire Cat'' galaxies presciently suggested to be responsible for much of the \lya forest by \citet{salpeter93} and \citet{charlton95}.

The number of \lya + metal-line absorbers produced by starburst winds could be even higher if more luminous galaxies also contribute to this process. While there is no compelling evidence for this at low-$z$, at $z \sim 3$, \citet{adelberger03} find a very strong correlation between \ion{C}{4} absorbers and Lyman break galaxies (LBGs), that may be evidence of massive galaxies expelling gas into the IGM at early epochs \citep{pettini01}. However, the galaxy population represented by the LBGs is unclear; they may be galaxies with stellar masses comparable to L$^*$ galaxies today or they may be dwarf galaxies undergoing intense bursts of star formation. Both of these models are possible given the observational constraints available. For example, \citet{papovich01} infer sub-0.1M$^*$ stellar masses for LBGs in the Hubble Deep Field, but \citet{shapley01, erb03}, using the \citet{steidel96a} LBG sample, infer significantly larger masses, up to the stellar mass expected for current L$^*$ galaxies. In addition, the sizes of these star forming regions is small, 2.5-5.2 kpc \citep{steidel96b}. However, both groups note that the presence or absence of an older stellar population is not well-constrained by the available data, so that these masses and sizes must be treated as lower limits for the galaxy as a whole. So, these LBGs can either be starbursting dwarfs that will assemble into larger galaxies at later epochs or the UV-luminous star forming regions inside more massive galaxies already assembled at the epoch of observation. If the former is the case, then a consistent picture emerges for both high and low redshift in which dwarf galaxies are primarily responsible for enriching the IGM with metals. If the latter is the case, then the spread of metals at early epochs is due to galaxies of a wider range of masses than in the present epoch and the estimate of the number of QSO metal absorption line systems made above is a lower limit.
 
Despite the huge number of metal-line systems which could be due to dwarf galaxy winds, these superwinds would not fill an appreciable fraction of the IGM. If the expansion speed associated with these outflows is similar to what we have found here, the volume filling factor of these winds decreases with cosmic time; i.e., superwind expansion speed $<$ Hubble expansion speed. Even if there were about ten times more dwarfs at high-z than now and they all produced starburst superwinds before merging into larger galaxies, the volume filling factor of metal-enriched gas is less than 1\%. This is consistent with numerical simulations of the spread of metals \citep{gnedin98} and suggests that the bulk of the \lya forest is unenriched. While \citet{gnedin98} and \citet{aguirre01} found that galaxy interactions could spread these metals a bit more widely, we would not expect that metals would be spread much further than $\sim 100$ kpc from galaxies today unless there were a very large number of star formation sites spread very uniformly throughout the Universe. \lya absorbers in regions most remote from galaxies, those in galaxy voids \citep[Paper III and][]{mclin02a}, are the best locations for testing this hypothesis. If early star formation were more widespread than is possible assuming all dwarfs create superwinds, even ``void absorbers'' would contain metals at abundances comparable to the possible ``floor'' of metallicity suggested by the recent work on low column density absorbers at high-$z$ \citep[${[{\rm Z}]} = -2.5$ Solar;][but see \citealp{simcoe04}]{cowie98}.

While the predicted number density of dwarf galaxy winds mentioned above suggests that virtually all \ion{C}{4} metal-line absorbers at high-$z$ could be due to this process, this certainly depends upon the detectability of \ion{C}{4} absorption in dwarf wind outflows. Indeed, the 3C~273 1586 \kms absorber does not contain detectable \ion{C}{4} absorption, being too dense for the current epoch ionizing background to produce significant amounts of triply ionized carbon (T2002). At $z \sim 2$ the ionizing background is $\sim 100$ times stronger and so an absorber of similar density would contain detectable \ion{C}{4}, because \ion{C}{4}/\ion{H}{1} is at a maximum for $\log{U} = 0$ to $-2$ and thus for physical densities of $10^{-2.5}$ to $10^{-4.5}$ cm$^{-3}$ at $z \sim 2$. Thus, assuming comparable physical size scales for absorbers in dwarf winds at $z \sim 2$ as at $z \sim 0$, \ion{C}{4} should be the dominant carbon species for absorbers with N$_{\rm H~I}= 10^{14}$ to $10^{16}$ cm$^{-2}$. And so \ion{C}{4} should be the most observable metal line in high-$z$ dwarf wind clumps which have already cooled sufficiently to be in photoionization equilibrium with the ionizing background at that epoch. Of course, \ion{C}{4} could also be produced by collisional ionization of these wind clumps with the IGM or with other pieces of the outflow. However, the cooling time at ${\rm T} \sim 10^5$ K is very short and would be unlikely to create wind clumps which are often detectable in \ion{C}{4} or even in \ion{O}{6}. For collisionally ionized gas, one would expect the column of \ion{O}{6} to be larger than for \ion{C}{4} \citep{heckman02}, and so more detectable. Absorption systems in which both \ion{C}{4} (and other lower ion species) and \ion{O}{6} are detected are thought to be multi-phase gas with both photo- and collisionally-ionized gas present \citep{giroux94}.

Recent work on the so-called ``weak \ion{Mg}{2}'' absorbers by \citet{churchill99}, \citet{rigby02}, and \citet{charlton02} have led these authors to speculate along similar lines to those presented here. Like the 1586 \kms absorber, standard photoionzation models of many of the weak \ion{Mg}{2} systems yield absorber thicknesses of $<$ 1 kpc. For the 1586 \kms absorber, a small size is also obtained and \ion{Mg}{2} absorption is also weak. Using the photoionization model of T2002 and a Solar ratio of Mg to C, we would predict an equivalent width of \ion{Mg}{2} 2798\AA\ of 10-20 m\AA, less than the limit (45 m\AA) we have derived from the available GHRS data (see Section 2). Thus, the absorber studied here is a ``local'' example of a ``weak \ion{Mg}{2}'' absorber. \citet{charlton02} interpret the small sizes of ``weak \ion{Mg}{2}'' absorbers as indicative of their three dimensional diameters, which require a huge number of these tiny spheres to create the substantial absorber numbers seen. Here, instead, we have suggested a spherical geometry of an expanding shell, with a large cross section on the sky but with small thicknesses along the line-of-sight, as inferred from photoionzation modeling. The substantial number of tiny spheres envisioned by \citet{charlton02}, several hundreds to thousands per L$^*$ galaxy, makes it difficult to imagine their origin except in some unknown population of objects. But an expanding supershell surrounding a fading dwarf galaxy could be viewed as tens to hundreds of individual filaments (or tiny spheres) as is commonly observed for individual supernova remnants in our own and nearby galaxies. Thus, we suggest that these two views are compatible and the differences largely semantic, so that no novel population of objects is required to explain their numbers.

However, a substantial percentage (25\%) of the weak \ion{Mg}{2} absorber population have been found by \citet{rigby02} to have Fe abundances comparable to Solar metallicity. This would be very unexpected for any absorber produced by outflowing winds from a dwarf galaxy, both given the low metallicity of stars and gas seen in dwarfs \citep[e.g., I Zw 18;][]{izotov99} and the substantial amount of even lower metallicity gas that an outflowing wind might be expected to sweep up in its progress through the IGM. If these high metallicities are correct, they argue for large numbers of undetected star forming pockets outside of galaxies, and against the dwarf galaxy wind model proposed here. These as-yet undetected star formation sites also would be the high-$z$ sources of metals required to produce a universal ``floor'' of metallicity \citep[see][]{aguirre01}. However, the method used by \citet{rigby02} to determine metallicity makes use of metal line widths and the \ion{H}{1} \lya profile only. Without, for example, information about the higher order Lyman series absorption lines, this method may not be accurate. Recently, \citet{shull00} have used UV spectra of bright AGN to show that the curve-of-growth method using the Lyman lines observed by FUSE obtains $b$-values which are a factor of two smaller in the mean than the $b$-values obtained from \lya profile fitting alone. This difference leads to an upward correction in hydrogen column density and thus to a downward correction in metallicity. Because the 1586 \kms absorber has seven detected Lyman lines \citep{sembach01}, the hydrogen column density derived by this method is quite secure. As a result, this absorber can provide a test of the accuracy of the \citet{rigby02} method for determining metallicities. Following their method, we begin with the equivalent widths and $b$-values measured for well-detected metal lines (see T2002). We then constructed a grid of photoionization models, requiring that the models be compatible with all metal line ratios, widths and limits. This procedure sets a limit on the temperature of the gas, which we can use to constrain the \ion{H}{1} line width associated with the metal-bearing gas. Fixing the \lya redshift and line width to the values required for the metal-bearing gas, and allowing additional \lya components, we then find the \ion{H}{1} column density associated with the metals. Using this procedure on the metal-line and \lya data for the 1586 \kms absorber yields N$_{\rm H~I} < 10^{15}$ cm$^{-2}$ and a metallicity of $> 40\%$ Solar. This should be compared with the 6\% obtained for this absorber by T2002. Thus, Rigby et al. would have inferred a much higher metallicity for the 1586 \kms absorber by using their method.

Therefore, we assert that the metal abundances in the absorbers Rigby et al. have studied are uncertain and may be substantially less than they calculate. If this is correct, the dwarf superwind hypothesis is an excellent fit to the various types of ``weak \ion{Mg}{2}'' systems they have studied. And there is no requirement for as yet unobserved, very numerous, small pockets of star formation well outside galaxies. The variations in observed species and ionization parameter seen in weak \ion{Mg}{2} systems and other weak metal line systems seem explicable given the possible variations in sightlines through supernova remnants, which might intercept cool filaments or hotter intra-filament regions or both. High quality UV spectra of background sources viewed through Galactic supernova remnants are needed to test this hypothesis.

\section{Conclusions}

In this Paper we have reported the discovery of a dwarf, post-starburst galaxy 71h$^{-1}_{70}$ kpc away on the sky and at the same recession velocity to within the combined errors as the highest column density \lya absorber in the 3C~273 UV spectrum obtained with HST and FOS, GHRS and STIS. A detailed study of this absorber using a STIS E140M spectrum \citep{heap02} in conjunction with FUSE observations of higher order Lyman lines \citep{sembach01} has been made by T2002. Based upon standard photoionization models, T2002 found a metallicity for this N$_{\rm H~I}$ = 10$^{15.85}$ cm$^{-2}$ absorber of 6\% Solar, an absorber thickness along the line-of-sight of 70 parsecs and an overabundance of Si to C of 0.2 dex relative to Solar. Using Lick absorption line indices measured from a ground-based spectrum of the nearby dwarf (M$_B = -14$) obtained at the Apache Point Observatory's 3.5m telescope, we estimate that the galaxy's stellar population has a mean age of 3.5 $\pm$ 1.5 Gyrs and a mean metallicity of 3-30\% Solar. Strong Balmer absorption (e.g., 2 \AA\ equivalent width of H$\delta$) and the absence of [O II] and H$\alpha$ emission, as well as the absence of \ion{H}{1} 21 cm emission (M$_{\rm H~I} \leq 5 \times 10^6$ M$_{\Sun}$) are strongly suggestive, but cannot prove, that this galaxy experienced a massive starburst $\sim$3.5 Gyrs ago, whose subsequent supernovae blew out all of its remaining gas. Given this interpretation, the evidence for a direct connection between this small galaxy and the nearby \lya absorber is quite convincing: approximate matches in recession velocity and metallicity, and a Si/C ratio in the absorber suggesting recent supernova type II enrichment. Despite an extremely thorough search for faint galaxies near the 3C~273 sightline \citep[M93;][and the present work]{rauch96,vangorkom93}, only this one dwarf has been found close to this absorber. The next nearest galaxies are two other dwarf galaxies $\approx 250$h$_{70}^{-1}$ kpc away, so that the only viable alternative picture for the 3C~273 absorber is gas produced by at least several nearby galaxies, including this dwarf.

Optical (B,R) images of this dwarf obtained at the APO 3.5m show that, despite the galaxy's elliptical appearance, its surface brightness profile is much better fit by an exponential disk with scale length 0.4h$^{-1}_{70}$ kpc and no optical peculiarities in the B-band image that would indicate the presence of dust. If a massive starburst did occur in this galaxy $\sim 3.5$ Gyrs ago, the type II supernovae would have had sufficient energy to accelerate several $\times 10^8$ M$_{\Sun}$ of \ion{H}{1} to the speeds necessary to reach 70h$^{-1}_{70}$ kpc radius by now. Thus, although we cannot prove beyond all doubt that the atoms producing the absorber came from this dwarf galaxy due to a massive starburst superwind, this seems to be the most logical conclusion. Also, since spectral synthesis models of the evolving stellar population of starbursts \citep{bruzual93} suggest that this galaxy was $\sim 10$ times brighter in B-band during its starburst phase and will fade ten times more in the next few Gyrs, this galaxy is an example of the intermediate stage in the evolutionary process that \citet{babul92} suggests begins as a blue, starbursting dwarf galaxy and ends as a dormant, dwarf spheroidal.

The primary uncertainty in this proposed scenario is that this galaxy experienced a single, massive starburst $\sim$3.5 Gyrs ago, which accounts for its current luminosity and stellar properties. A starburst of this magnitude in a dwarf galaxy \citep[requiring 1 M$_{\Sun}$ yr$^{-1}$ for 10$^8$ yrs;][]{martin03} is unexpectedly large. Further, the indirect evidence from observations of the stellar populations of Local Group dwarf spheroidals is that low luminosity galaxies typically experience several epochs of star formation \citep{mateo98}. However, even in the case that the starburst which took place in this galaxy accounts for only $\sim 30\%$ of all of its stars and occurred as recently as 2 Gyrs ago, a supernovae-driven wind can still account for the observed properties and location of the 1586 \kms absorber towards 3C~273. A near-UV image of this dwarf galaxy could determine whether the massive, single starburst model is correct. \citet{bruzual93} models of instanteous starbursts have little mid-UV flux more than 1 Gyr after the starburst, fading from 6 to 20 times less flux than at B-band between 1 and 4 Gyrs after the starburst. However, stellar synthesis models with several star formation epochs or continuous star formation have sufficient blue horizontal branch stars so that the mid-UV flux never gets this low compared to B-band \citep[see e.g., \citet{brown97} on the ``UV-upturn'' in ellipticals, which has an onset of $\geq 7$ Gyrs after a starburst;][]{tantalo96}.

While the association of this dwarf galaxy with a \lya + metal-line absorption system is interesting as an individual example of a ``Cheshire Cat'' galaxy \citep{salpeter93,charlton95}, it also may be indicative of a large population of starbursting dwarf galaxies, whose supernovae can drive gas away from these dwarfs due to their shallow gravitational potential wells \citep{maclow99,ferrera00}. We are currently searching for a faint galaxy near the \RXJ\ sightline that could be responsible for a very similar pair of metal-line absorbers at 1666 \kms, $\sim 350$h$^{-1}_{70}$ kpc away on the sky from the 1586 \kms absorber in 3C~273. If a dwarf galaxy is present near \RXJ\, this will provide further proof in favor of the dwarf starburst superwind model for these absorbers. All other known metal line systems are too far away ($cz \geq 3000$) to allow dwarfs similar to this one to be observed by current surveys for galaxies near QSO absorbers (i.e., at m$_B > 19$; M93). Therefore, metal-line absorbers would be identified with more luminous galaxies at greater impact parameters \citep[e.g.,][]{chen01,adelberger03}, which happen to be in the same galaxy filament as a dwarf (or dwarves) closer to the sightline.

We view the discovery of this dwarf close to a \lya + metal-line absorption system as strong evidence in favor of the \citet{babul92} scenario used to explain the large population of ``faint, blue galaxies'' seen on deep optical images. The large numbers of faint, blue galaxies plus the theoretical expectations of standard heirarchical pictures of galaxy formation strongly suggest that there were many more of these dwarf galaxies in the past than now. If most of these dwarfs went through a violent starburst phase, dwarf starburst superwinds could be responsible for ${\rm dN/dz} \gtrsim 500$ \lya + metal-line absorption systems in the spectra of high-$z$ QSOs. This estimate is based upon the assumption that all dwarf galaxies produce a 75h$^{-1}_{70}$ kpc radius shell of metal-enriched gas around them before they  merge to form larger galaxies. Evidence from other work \citep[e.g., ][]{bergeron91,steidel95,steidel98} is that larger galaxies (${\rm L} > 0.1$L$^*$) form bound halos of denser gas and may not contribute significantly to the spread of metals into the IGM. But, even if all dwarfs produce massive starburst superwinds, $< 1\%$ of the IGM would become enriched in metals, leaving most of the \lya forest absorbers metal-free at the lowest column densities. This conclusion can be tested by determining if \lya absorbers in galaxy voids \citep[e.g., ][]{mclin02a} contain metals at a level $\geq 0.3\%$ Solar.

We thank Erica Ellingson and the APO 3.5m observing specialists for assistance in obtaining the dwarf galaxy spectrum. We thank Carnegie Observatories for their support of the wide field CCD multi-object spectroscopy system and the WIYN queue observers for their excellent observing support. We thank Bianca Poggianti for the use of her figure from a previous publication, Yeong-Shang Loh for accessing the public Sloan Digital Sky Survey database to provide information for us on the galaxies near the 3C~273 sightline and Simon Morris for helpful discussions of his galaxy survey work near 3C~273. JTS, KMM, BAK and JLR acknowledge support from NASA HST General Observer grants GO-06593.01-A, GO-08182.01-A, GO-09506.01-A and archival research grant AR-09221.01-A. JLR also acknowledges support from an NSF Astronomy \& Astrophysics Postdoctoral Fellowship, AST-0302049.

\clearpage

\clearpage
\begin{deluxetable}{cc}

\tabletypesize{\scriptsize}
\tablecaption{Measured Lick Indices \& Emission Line Limits \label{tbl-lick}}
\tablewidth{0pt}

\tablehead{\colhead{Lick Index Name} & \colhead{Measured Value in Dwarf}}

\startdata
H$\delta_F$ & $1.9\pm0.7$ \AA \\
H$\gamma_F$ & $1.9\pm0.9$ \AA \\
H$_{\beta}$ & $3.7\pm0.9$ \AA \\
Mg$_2$      & $0.08\pm0.02$ mag \\
Fe5270      & $1.2\pm0.8$ \AA \\
Fe5335      & $1.2\pm0.7$ \AA \\
$<$Fe$>$    & $1.2\pm0.5$ \AA \\
\\
\multicolumn{2}{c}{\it Equivalent Width Limits} \\
\tableline \\
$[$OII$]$ $\lambda$3727  & $< 5$ \AA \\
$[$OIII$]$ $\lambda$5007 & $< 0.6$ \AA \\
H~$\alpha$               & $< 0.6$ \AA \\
\enddata

\end{deluxetable}

\begin{deluxetable}{cc}

\tabletypesize{\scriptsize}
\tablecaption{Absorber/Galaxy Connections \label{tbl-compare}}
\tablewidth{0pt}

\tablehead{\colhead{3C~273 Absorber} & \colhead{Dwarf Galaxy}}

\startdata
$cz=1586\pm5$\kms & $cz=1635\pm50$\kms \\
log N$_{\rm H~I} = 15.85$ cm$^{-2}$ & $b = 71{\rm h}_{70}^{-1}$ kpc \\
log n$ = -2.85$ cm$^{-3}$ & ${\rm m}_B = 17.9 \; ({\rm M}_B = -13.9$) \\
log U$ = -3.0$ & log$(L_B/L_{\Sun}) = 7.8$ \\
Shell thickness$ = 70$ pc & (B$-$R)$ = 1.2$ \\
$[$Fe/H$] = -1.2$ & M$_{\rm H~I} < 3 \times 10^6$ M$_{\Sun}$ \\
$[$Si/C$] = 0.2$ & $[$Fe/H$] = -1.1$ \\
& Mean Stellar Age$ = 2$-3 Gyr \\
\enddata

\end{deluxetable}

\end{document}